\documentclass[runningheads]{llncs}

\usepackage[utf8]{inputenc}
\usepackage{graphicx}
\usepackage{url}
\usepackage{hyperref}
\usepackage{xcolor}

\hypersetup{
    colorlinks=true,                
    breaklinks=true,                
    urlcolor= blue,                
    linkcolor= blue,                
    bookmarksopen=false,
    filecolor=black,
    citecolor=blue,
    linkbordercolor=blue
}

\begin{document}

\title{Executable Models and Instance Tracking for Decentralized Applications on Blockchains and Cloud Platforms - Metamodel and Implementation}

\titlerunning{Executable Models and Instance Tracking for Decentralized Applications}

\author{Felix Härer\orcidID{0000-0002-2768-2342}} %\Envelope

\authorrunning{F. Härer}

\institute{Digitalization and Information Systems Group, University of Fribourg, Switzerland
\email{felix.haerer@unifr.ch}\\
\url{https://www.unifr.ch/inf/digits/}}

\maketitle

\begin{abstract}
Decentralized applications rely on non-centralized technical infrastructures and coordination principles. Without trusted third parties, their execution is not controlled by entities exercising centralized coordination but is instead realized through technologies supporting distribution such as blockchains and serverless computing. Executing decentralized applications with these technologies, however, is challenging due to the limited transparency and insight in the execution, especially when involving centralized cloud platforms. This paper extends an approach for execution and instance tracking on blockchains and cloud platforms permitting distributed parties to observe the instances and states of executable models. The approach is extended with (1.) a metamodel describing the concepts for instance tracking on cloud platforms independent of concrete models or implementation, (2.) a multidimensional data model realizing the concepts accordingly, permitting the verifiable storage, tracking, and analysis of execution states for distributed parties, and (3.) an implementation on the Ethereum blockchain and Amazon Web Services (AWS) using state machine models. Towards supporting decentralized applications with high scalability and distribution requirements, the approach establishes a consistent view on instances for distributed parties to track and analyze the execution along multiple dimensions such as specific clients and execution engines.


\end{abstract}

\section{Introduction}

The execution of applications has evolved over the years towards further distribution, involving a recent shift from traditional client-server to advanced cloud-based architectures~\cite{liServerlessComputingSurvey2022}. Given this trend, applications are deployed increasingly on cloud platforms that support their execution at scale while allowing for distribution across servers, e.g. in multiple geographic locations. In parallel to these developments, decentralized applications based on blockchains~\cite{wuFirstLookBlockchainbased2021} have emerged where (1.) the deployment and execution do not depend on centralized technical infrastructures, e.g. individual servers, and (2.) the parties involved in deployment, execution, or any governance aspects, coordinate their actions without centralized control~\cite{harerExecutableModelsInstance2022a}, e.g. using roles, permissions, and voting mechanisms instead of relying on central authorities or trusted third parties. Decentralized applications can be applied for decentralized organizations~\cite{wangDecentralizedAutonomousOrganizations2019,harerIntegrierteEntwicklungUnd2019}, processes~\cite{harerProcessModelingDecentralized2020}, industrial applications~\cite{caiDecentralizedApplicationsBlockchainEmpowered2018} such as IoT, supply chain management, source tracing, and further applications based on credits or tokens~\cite{caiDecentralizedApplicationsBlockchainEmpowered2018}. Blockchains provide verifiable execution in these scenarios, i.e., all parties involved are able to transparently observe individual transactions and executions in smart contracts.

For executing decentralized applications, cloud platforms are currently limited as they support decentralized technical infrastructures (1.) and decentralized coordination (2.) only partially. Decentralized infrastructures can be utilized with serverless computing~\cite{liServerlessComputingSurvey2022}, where execution is specified abstract from individual servers or virtual machines and is distributed and scaled by the cloud platform. When involving distributed parties without centralized coordination, however, it becomes challenging for them to coordinate their actions due to the limited insight in the execution on cloud platforms. Here, the execution is controlled in a centralized manner while distributed parties cannot observe and follow it over time. Blockchain-based execution, on the other hand, allows for transparency in this regard but is limited in scalability~\cite{ZhouScalability2020} required for performing data- and compute-intensive operations. Thus, an execution on cloud platforms that utilizes blockchains to gain insight in the execution is investigated in this paper, motivated by the overarching research question: "How can distributed parties track the execution behavior of applications on cloud platforms and blockchains without centralized coordination?".

% In particular, it is not obvious how distributed parties might construct a consistent representation of the execution, distribute it, and analyze it in order to verify and control execution behavior. 

Towards addressing this question, an approach for execution and instance tracking is applied~\cite{harer_decentralized_2018,harerExecutableModelsInstance2022a} that tracks the instance of each execution by recording individual instance states of an executable model in a distributed fashion. When running an executable model on a cloud platform, e.g. a process, workflow, or state machine model, instances with their states are registered in a smart contract that allows distributed parties to observe execution behavior.

The contribution of this paper is an extension of the architecture for executable models and instance tracking on blockchain and cloud platforms~\cite{harerExecutableModelsInstance2022a} by applying conceptual modeling and metamodeling~\cite{karagiannisDomainSpecificConceptualModeling2022,karagiannisMetamodelingApplicationAreas2008} together with multidimensional data modeling~\cite{gosainConceptualMultidimensionalModeling2015} towards an implementation. In particular, the contribution encompasses (1.) a metamodel for tracking instances among distributed parties independent of concrete models or implementation, (2.) a multidimensional data model for recording instances locally at each participant providing verifiable storage, the tracking of execution behavior in instance states, and analysis of past executions along multiple dimensions such as time or clients, as well as (3.) an implementation demonstrating feasibility on the Ethereum blockchain~\cite{buterin_ethereum:_2013} and Amazon Web Services (AWS) using state machine models in the form of AWS Step Functions~\cite{amazon_2023}. 

The remainder of this paper is structured as follows. Section~\ref{background-related-work} introduces background on relevant technologies and related work on model-based approaches utilizing blockchains. Section~\ref{metamodel-datamodel} describes the design of the metamodel, applies it for the design of a multidimensional data model, and leads to Section~\ref{implementation} with an implementation for demonstrating feasibility. Section~\ref{discussion-outlook} discusses and concludes the paper.

\section{Background and Related Work}
\label{background-related-work}

This section introduces background for blockchains, serverless computing, and model-based representations, and discusses related work utilizing models on blockchain infrastructures.

Blockchain technologies permit distributed parties to verifiably store data in integrity-secured data structures and to verifiably compute using programs in the form of smart contracts. Commonly, blockchain systems can be described in terms of a data structure of transactions, e.g. backward-linked blocks~\cite{nakamotoBitcoinPeertoPeerElectronic2008,buterin_ethereum:_2013} or graphs~\cite{wangSoKDAGbasedBlockchain2023}, a consensus algorithm executing and verifying transactions~\cite{bouragaTaxonomyBlockchainConsensus2021}, and a network of distributed parties operating blockchain nodes for distribution and verification of the data structure as defined by the consensus algorithm~\cite{dotanSurveyBlockchainNetworking2021}. In contrast to cloud platforms, the execution in open and permissionless blockchains such as Ethereum or Cardano~\cite{harerInteroperabilityOpenPermissionless2022a} is transparent, verifiable by all distributed parties without trusted third parties, and especially suited for decentralized applications. Here, decentralized applications~\cite{wuFirstLookBlockchainbased2021} can utilize smart contracts for verifiable computation to achieve (1.) execution without relying on centralized technical infrastructure and (2.) non-centralized coordination of the execution controlled and observed by distributed parties~\cite{harerExecutableModelsInstance2022a}. Cloud platforms such as Amazon Web Services (AWS), Microsoft Azure, or Google Cloud Platform focus on providing infrastructure, software platforms, or applications as a service~\cite{nadeemEvaluatingRankingCloud2022}. While infrastructure services commonly concern specific physical or virtual servers, the services for platforms and applications hide infrastructure such that, possibly, non-centralized infrastructure might be used under the control of the provider. This is supported by the increasing abstraction recently observed in platform services, especially regarding serverless computing~\cite{hassanSurveyServerlessComputing2021}. Here, specifications tend to be modular and functional, e.g. using lambda functions, permitting distribution for availability and reliability as well as scalability in terms of parallel execution on distributed servers. In addition, model-based execution specifications can be applied, e.g., on Microsoft Azure or AWS~\cite{curty_blockchain_2022}.

% For data analytics and machine learning~\cite{SAGE MAKER}, 3D models for augmented and virtual reality, and state machines for distributed applications.

Model-based approaches utilizing blockchains as technical infrastructures have been suggested, e.g., to store, track, or execute models with their instances. Related works discuss often-times the execution of business processes and workflows~\cite{weberUntrustedBusinessProcess2016,garcia-garcia_using_2020}, e.g. using the BPMN modeling language~\cite{omgBusinessProcessModel2014} for processes~\cite{lopez-pintadoCaterpillarBusinessProcess2019,curty_blockchain_2022} or choreographies~\cite{ladleifModelingEnforcingBlockchainBased2019b} and related modeling languages such as CMMN~\cite{milaniModellingBlockchainbasedBusiness2021}, with further approaches discussing, e.g., ontologies for execution~\cite{choudhuryAutoGenerationSmartContracts2018c}, state machine execution~\cite{nakamuraInterorganizationalBusinessProcesses2018,mavridouDesigningSecureEthereum2018}, enterprise model storage~\cite{fillKnowledgeBlockchainsApplying2018a}, attestation of models~\cite{harerDecentralizedAttestationConceptual2019b,harerDecentralizedAttestationDistribution2022a}, as well as instance tracking~\cite{harer_decentralized_2018,harerExecutableModelsInstance2022a} and the design of process models and business systems~\cite{harerProcessModelingDecentralized2020}. The attestation and instance tracking approaches are applied in this paper to capture the states of model instances over time and utilize the blockchain for distributing the metadata of models such that distributed parties can observe and verify the progression of instance states. For the execution of decentralized applications, these approaches can be combined with serverless computing that supports execution abstract from technical infrastructure and offers scalability. Towards this direction, the previously introduced architecture for blockchains and cloud platforms~\cite{harerExecutableModelsInstance2022a} is used as a basis. Components of the architecture are clients, controlling and tracking the execution of models and instances, cloud platforms with execution engines, and blockchains with a smart contract to distribute and register models, instances, and states. 
 
\section{Instance Tracking Metamodel and Data Model}
\label{metamodel-datamodel}

The two following subsections describe the metamodel in its concepts and subsequently apply it for the design of a multidimensional data model to track instances and to analyze past executions.

\subsection{Metamodel}
\label{metamodel}

\begin{figure}[htb]
  \centering
  \includegraphics[clip, trim=0.1cm 0cm 0.01cm 0.01cm, width=1\linewidth]{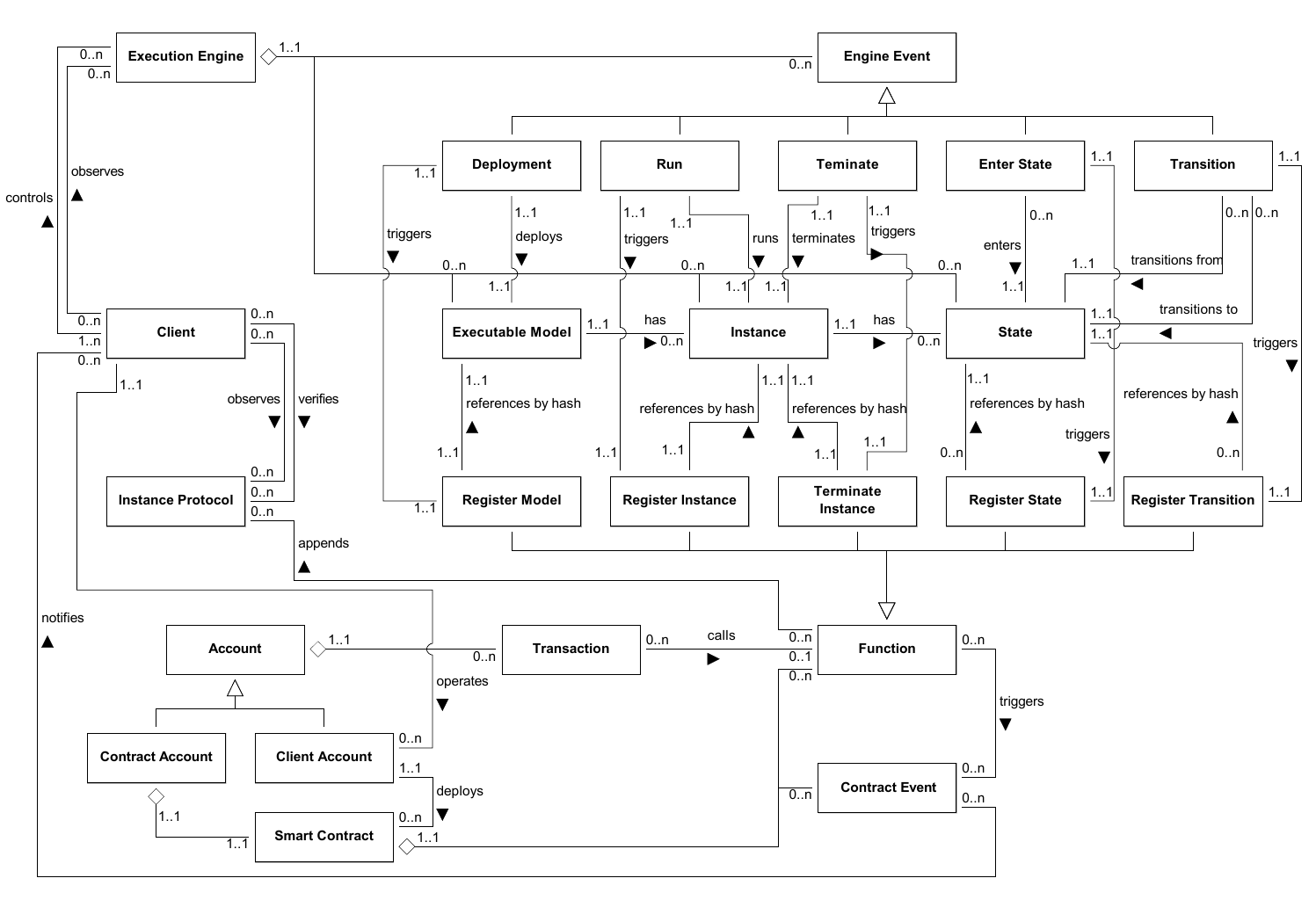}
  \caption{Metamodel describing the concepts for model-based execution and instance tracking. Clients exercise control over an execution and observe models with instances and states through the events of a smart contract. }
  \label{fig:metamodel}
  \vspace{-1mm}
\end{figure}

Metamodeling~\cite{karagiannisDomainSpecificConceptualModeling2022,karagiannisMetamodelingApplicationAreas2008} is applied here for the identification and description of concepts and relationships required for executable models and instance tracking on blockchains and cloud platforms. Based on the previously introduced architecture~\cite{harerExecutableModelsInstance2022a}, the metamodel in Figure~\ref{fig:metamodel} explicitly describes and generalizes the concepts required for execution engines, smart contracts, and clients controlling and observing the execution behavior. In the following, the related metamodel concepts are discussed from a client point-of-view.

\subsubsection{Client and Execution Engine Concepts}
\label{client-exec-concepts}

A \texttt{Client} represents one of the distributed parties involved in an execution. Clients can exercise control over execution engines, observe instance states, and verify the states with instance protocols obtained through interactions on the blockchain.

An \texttt{Execution Engine} supports running executable models under the control of a client. Examples include cloud platforms such as Amazon Web Services (AWS) that support, e.g. workflow models for orchestrating web services, low-code and no-code platforms such as OutSystems~\cite{curty_blockchain_2022} that represent application logic and user interface flows visually, or workflow engines such as Camunda for executable business process models~\cite{camundaCamundaBPMManual2022}. In operation, the lifecycle of an execution can be observed by following each \texttt{Engine Event}. Engine events encompass the \texttt{Deployment} of an executable model, \texttt{Run} a model by creating an instance, \texttt{Enter State} for each state reached over time, \texttt{Transition} for each transition between states, and \texttt{Terminate} for ending the execution of an instance. When running a deployed model, the engine might emit events for entering states and possibly events for transitions. E.g., AWS Step Function models emit events when entering a state and when initial and final states of an execution are reached~\cite{amazon_2023}. 

An \texttt{Executable Model} is a representation that specifies execution abstract from source code. Executable models are represented in a custom modeling language specified by the execution engine or using a standardized modeling language. For instance, AWS Step Function models utilize a custom Amazon States Language~\cite{amazon_2023} while process and workflow models might rely on the BPMN 2.0 standard~\cite{omgBusinessProcessModel2014}.

An \texttt{Instance} is created when a deployed model is run. Each instance contains
concrete data or objects for the abstract concepts defined in the model as far as they are related to the individual execution. An engine such as the AWS engine for Step Functions might issue an instance ID and run the model in a cloud-based execution environment, thereby entering an initial state. 

A \texttt{State} reached during the execution describes a discrete state an instance has at a point in time. When observing the execution from the point-of-view of a client, the result is a series of states for an instance, regardless of weather the engine emits events when entering a state or when transitioning between states. An instance might be terminated, e.g. when reaching a final state, in exception or failure cases, or when manually terminating the execution.

\subsubsection{Client and Blockchain Concepts}

For engine events to be distributed and observable by other clients, a smart contract deployed in a blockchain is utilized. 

A \texttt{Client} interacts with the blockchain through a \texttt{Client Account}, i.e., an account capable of deploying smart contracts and calling its functions by a \texttt{Transaction} that is signed by the \texttt{Client}. For this reason, a client exercising control over an execution engine deploys the instance tracking \texttt{Smart Contract} that is represented by a \texttt{Contract Account}. Thereby, the client links the execution engine under its control to the \texttt{Client Account}. This link can be observed together with future actions controlling the execution through the \texttt{Contract Account} by other clients. 

In particular, the client invokes the smart contract by a \texttt{Transaction} with a \texttt{Function} that is triggered for each \texttt{Engine Event}, linking event occurrences to a \texttt{Client Account}. For registering the engine events, the smart contract functions emit a \texttt{Contract Event} such that other clients will be notified and can reconstruct the instance state. Engine events of the smart contract are \texttt{Register Model} for deployment, \texttt{Register Instance} when a model is run, \texttt{Register State} when entering a state, \texttt{Register Transition} for transition events, and \texttt{Terminate Instance} upon the termination of an instance. For securing integrity, i.e. for verifying models, instances, and states have not been changed, each function records a hash value as a content-based representation of the model, instance, or state. For example, when a state is entered, the client controlling the execution obtains the state, calculates its hash value, and includes it in a smart contract transaction that calls the register state function. 

% E.g., with an API of web3.js https://web3js.readthedocs.io/en/v1.8.2/
% Events and functions are present and accessible through AWS APIs and the AWS Management Console for control from the client side

For tracking instances, a \texttt{Client} observes each \texttt{State} of an instance on the execution engine and verifies its existence and integrity with the smart contract in the following manner. Each \texttt{Engine Event} is reflected by a \texttt{Transaction} that calls the corresponding smart contract \texttt{Function} and triggers a \texttt{Contract Event}. Through the event, a \texttt{Client} is notified and appends the information of the event to an \texttt{Instance Protocol}, including the information of the function call. Thus, the protocol is created or reconstructed entirely from blockchain data.

For verification, the observed state from the execution engine is applied to a hash function for comparing the resulting value with the hash value supplied by the initial function call. In case of matching hash values, the integrity of an \texttt{Executable Model}, \texttt{Instance}, or \texttt{State} resulting from an \texttt{Engine Event} is confirmed. Furthermore, it can be linked to an \texttt{Execution Engine} controlled by a \texttt{Client} and verified regarding further metadata such as timestamps. In case the verification fails at any point, e.g. hash values differ, validity cannot be established. Consequently, the observed \texttt{Executable Model}, \texttt{Instance}, or \texttt{State} is discarded as it cannot be attributed to the \texttt{Client} that is linked to the smart contract, e.g., because of a modification made on the cloud platform not initiated by the \texttt{Client}. 

\subsection{Multidimensional Data Model}

\begin{figure}[htb]
  \centering
  \includegraphics[clip, trim=0.4cm 0cm 0.2cm 0cm, width=1.0\linewidth]{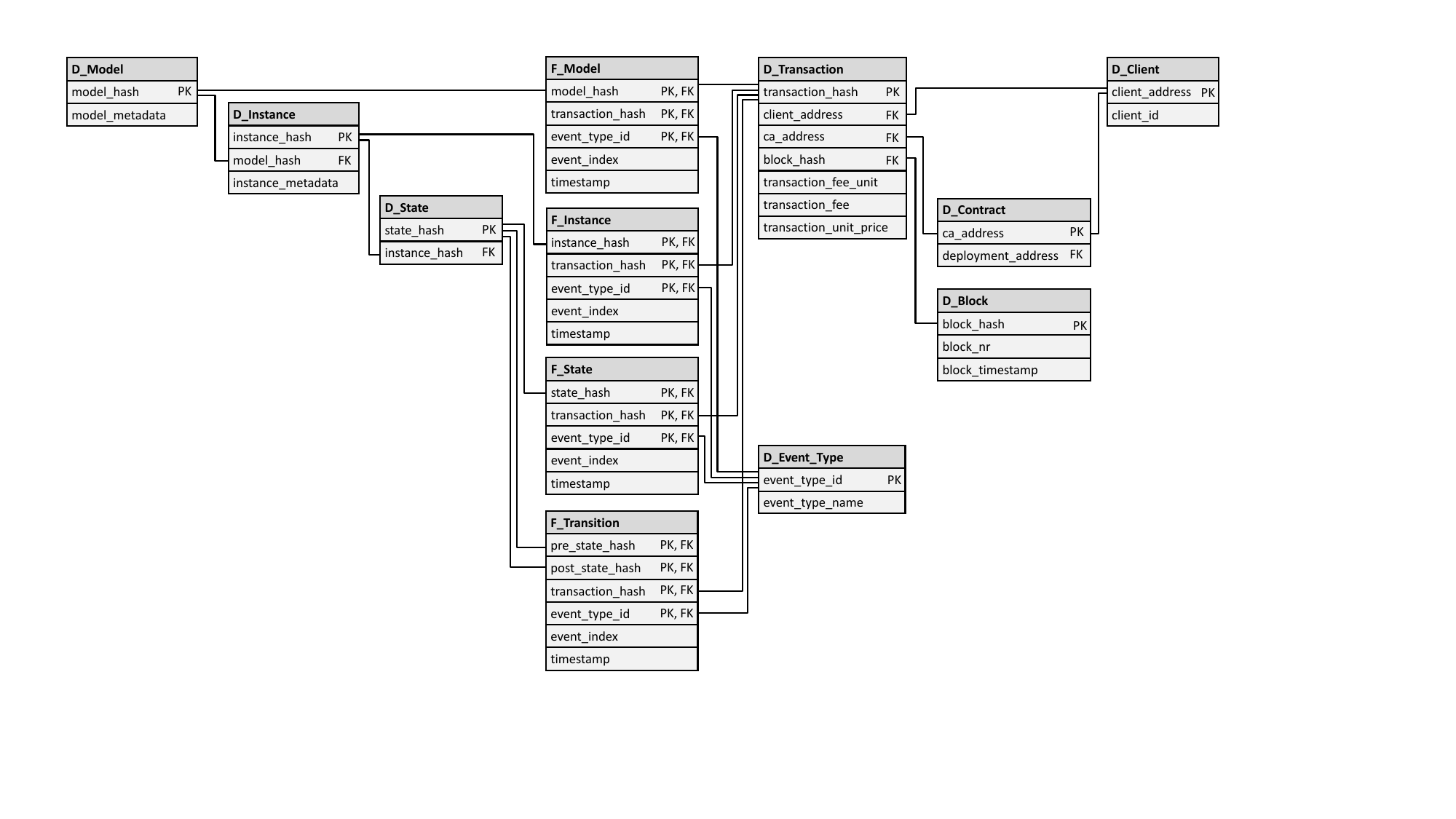}
  \caption{Multidimensional data model for tracking models, instances, states, and transitions according to smart contract events. Tracking and aggregation is supported over dimensional data (prefix D), e.g. for clients or contracts, for the events stored as fact data (prefix F). 
  }
  \label{fig:datamodel}
  \vspace{-4mm}
\end{figure}

On the basis of the metamodel, this section establishes a data model that can be used for tracking instances locally at each individual client. The data model design applies multidimensional modeling~\cite{gosainConceptualMultidimensionalModeling2015} to describe and analyze the data related to instances along multiple dimensions of models and their execution. Relational data is assumed for providing a data warehouse with a schema that can be efficiently analyzed with state-of-the-art databases. Figure~\ref{fig:datamodel} shows the data model. In the first of the two following subsections, the model is discussed for storing data along the dimensions, e.g. of states and their related events and transactions of the smart contract. The second subsection discusses the model for tracking and analyzing data by means of the facts in the form of events and timestamps.

\subsubsection{Dimensions}
\label{data-mod-dim}

Dimension tables store the data of models, instances, and states, as well as blockchain-related data of events and transactions between the client and the smart contract.

For models, instances, and states, the dimension tables \texttt{D\_Model}, \texttt{D\_Instance}, and \texttt{D\_State} store hash values for identification as primary key (PK). The hash values reference each model, instance, and state at a point in time that is captured by each value, calculated when an event is observed. Foreign key (FK) relationships indicate referential integrity such that states are recorded for a specific instance and instances are, in turn, recorded for a specific model. The content-based storage in the form of hash values also permits the verification of integrity, by comparing to values calculated from models, instances, and states on the execution engine.

Blockchain-related data is stored for every transaction between the client and the smart contract in \texttt{D\_Transaction}. A transaction is identified by a hash value (transaction\_hash) and indicates the client address, contract account address (ca\_address), the block of the transaction (block\_hash), and the transaction fee due to the execution on the blockchain. A fee is recorded with a unit (transaction\_fee\_unit), e.g. gas\footnote{\url{https://ethereum.org/en/developers/docs/gas/}}, and price per unit (transaction\_unit\_price). Blocks stored in \texttt{D\_Block} allow accessing a timestamp that can be used in verification by comparison to timestamps from the execution engine. In each transaction, the referenced contract account address (ca\_address) points to the contract in \texttt{D\_Contract} that, in turn, is linked to the specific client that initially deployed the contract (deployment\_address). With each client in \texttt{D\_Client} also an additional blockchain-external ID is stored in client\_id.

Event types of all observed events are recorded from the point-of-view of a client and recorded in \texttt{D\_Event\_Type}. Only the type is stored with an identifier (event\_type\_id), e.g. as generated by a database. According to the metamodel, event types observed at run-time are stored such that, e.g., the registration of models with the type "Register Model" is part of the dimensional data.

\subsubsection{Facts}
\label{data-mod-facts}

Fact tables store each occurrence of an event together with the related model, instance, state, or transition. Along the dimensions, events might be analyzed, e.g., for events of a specific type or from a particular client.

For events related to a model, \texttt{F\_Model} stores each occurrence in terms of the event type (event\_type\_id), the transaction the event was observed in (transaction\_hash), an index identifier within the transaction (event\_index), and the time the event was observed (timestamp). Similarly, events of instances, states, and transitions are stored in \texttt{F\_Instance}, \texttt{F\_State}, and \texttt{F\_Transition}. For each of the fact tables, the related models, instances, states, or transitions are referenced by hash values.  % (model\_hash)

Data in the fact tables reflect the models, instances, and states as they existed at the point in time when an event occurred. For tracking and the evaluation of past executions, a fact table joined with one or more dimension tables permits an analysis along the dimensions. E.g., a specific model, instance, or state might be analyzed over time, clients, contracts, transactions, blocks, or event types. In addition, data can be aggregated for the dimensions, e.g., accessing for all states of an instance, the number of states, timestamps, and duration of states, as well as statistics such as the total or average number of states in multiple instances, or the average state duration.

\section{Implementation}
\label{implementation}

The aim of this section is a demonstration of feasibility by implementation on the Ethereum blockchain and the AWS cloud platform using Step Function models. For this purpose, the architecture of the approach~\cite{harerExecutableModelsInstance2022a} has been implemented and tested with a prototype client application in Python 3.9 for event processing, a smart contract in Solidity 0.8.18 for registering and distributing events, and~a PostgreSQL 15 database implementation for storing instance protocols\footnote{\url{https://github.com/fhaer/Itrex-Engine-Event-Processing}}. The corresponding concepts of the metamodel are realized by tracking executable models, instances, and states in an event-driven fashion with local storage of instance protocols according to the data model. 

\subsection{Executable Models on AWS}

\begin{figure}[htb]
  \centering
  \includegraphics[clip, trim=0.0cm 0cm 0.01cm 0.01cm, width=0.9\linewidth]{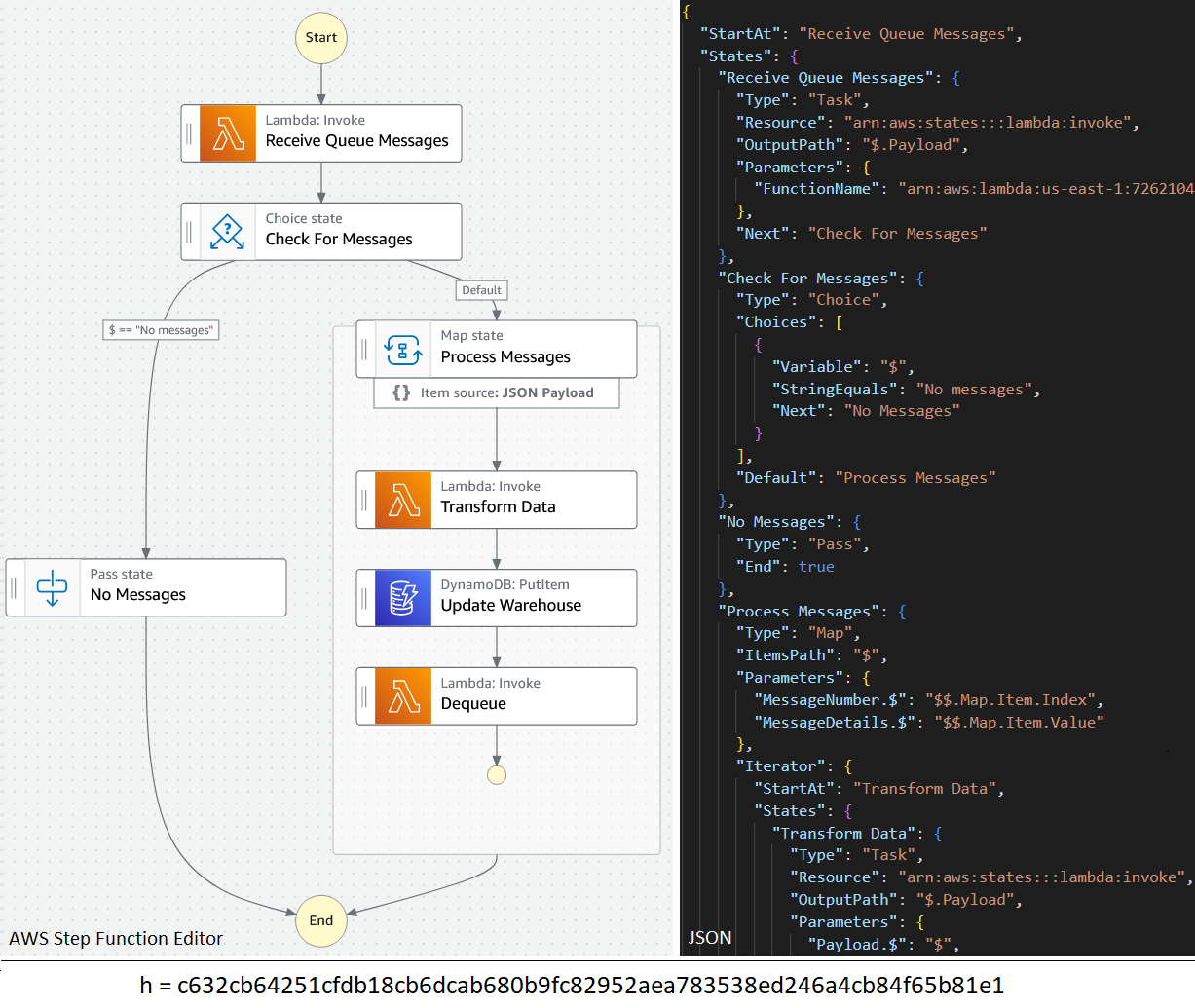}
  \caption{AWS Step Functions model for parallel data processing in a distributed scenario, where data is received through a message queue, transformed, and stored in a data warehouse. The model is shown as state machine diagram and JSON representation. A hash value $h$ is used for identification based on the model content and for verification. }
  \label{fig:aws-sf-m1}
  \vspace{-1mm}
\end{figure}

For the use case of parallel data processing in a distributed scenario, an example model has been implemented as shown in Figure~\ref{fig:aws-sf-m1} in a state machine diagram on the left-hand side and an excerpt of a corresponding JSON representation on the right-hand side. In this scenario, data records are received from a message queue, transformed, and written into a data warehouse. Such a pattern can be observed in many distributed applications, e.g., for compute- and data-intensive processing of business transactions, device data of industrial systems, IoT data, or analytics and machine learning. On cloud platforms, executable models might be composed out of services for these applications. In the example model, elements represent task states with service interactions in addition to initial and final states with further elements for mapping and control flow. Utilizing AWS services, messages are received using AWS Lambda with an SQS message queue and processed in parallel involving data transformation in Lambda, updating data warehouse records in DynamoDB, and dequeuing in Lambda. Further examples involve controlling other step function models in addition to, e.g., training machine learning models~\cite{amazon_2023}. 

\subsection{Prototype}

The client application operates in two modes, concerning (1.) event processing of AWS execution events with the processing of models, instances, and states, leading to smart contract function calls, as well as (2.) for tracking and analysis based on smart contract events to record models, instances, and states in an instance protocol with a data warehouse for further analysis. 

\subsubsection{Event Processing on AWS and Smart Contract Calls}

The client application observes AWS CloudWatch event logs for model deployments, running instances, entering states, and terminating instances. For each observed model, execution, and state, representations in JSON format are used, normalized regarding the formatting, and applied as input for calculating a hash value $h$ using SHA-256~\cite{nistSecureHashStandard2015} as content-based identifier. 

\begin{figure}[htb]
  \centering
  \includegraphics[clip, trim=0.4cm 0cm 0.01cm 0.01cm, width=1.0\linewidth]{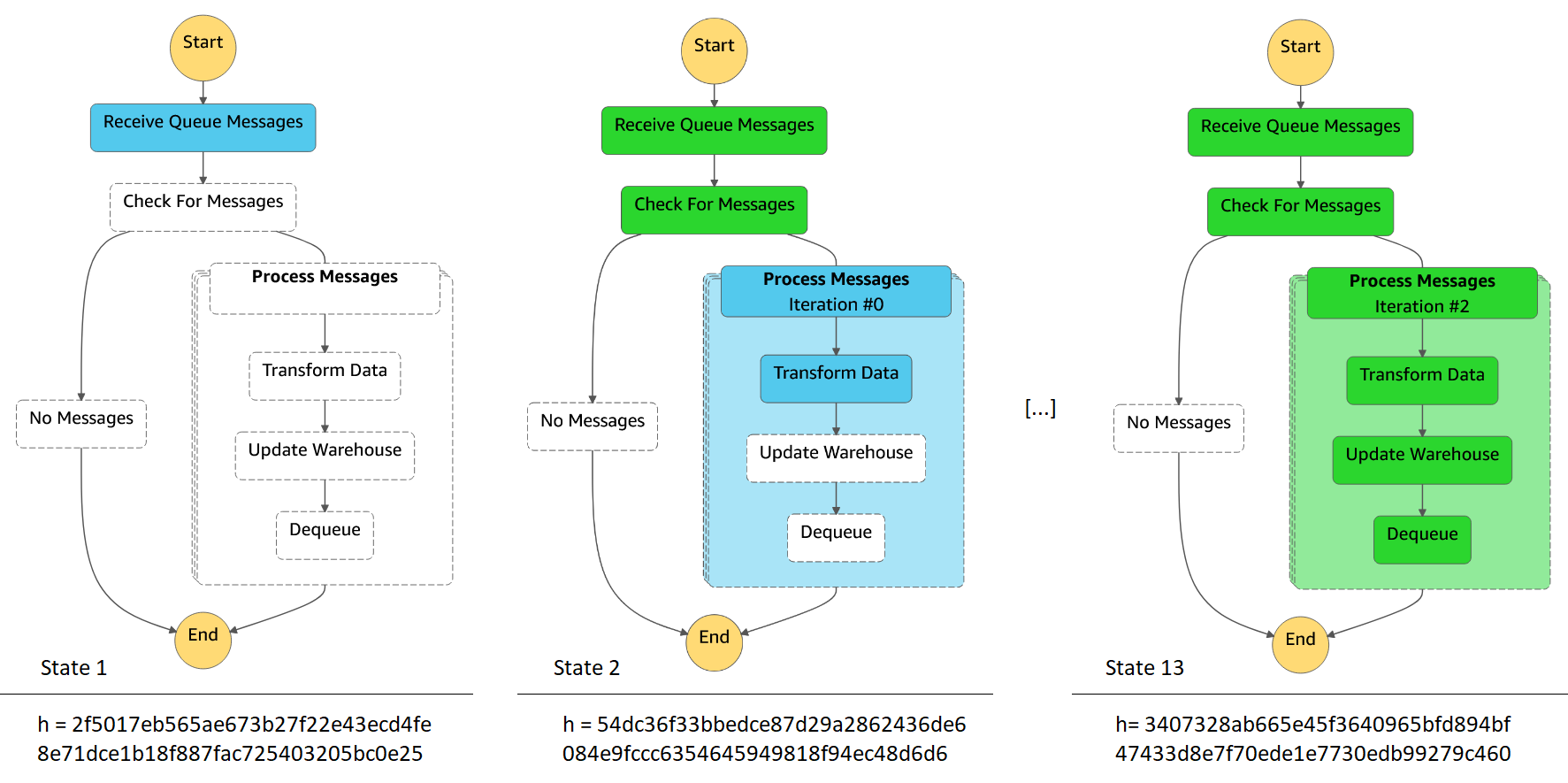}
  \caption{AWS Step Functions instance created from the model shown in Figure~\ref{fig:aws-sf-m1}. The instance consists of 13 states for parallel data processing, executed in multiple concurrent iterations. By chance, the last state of the instance belongs to "Iteration \#2". }
  \label{fig:aws-sf-m1-i1}
  \vspace{-1mm}
\end{figure}

With the deployment of a model, corresponding events are observed, the AWS JSON data is obtained, normalized, and represented in $h$, as indicated in the example at the bottom of Figure~\ref{fig:aws-sf-m1}. Subsequent events of instance runs and terminations are collected and applied to the hash function, based on JSON data created with a globally unique identifier (GUID), timestamps, and the AWS region. By this data, the start and end of each instance is represented. For the states of an instance, Figure~\ref{fig:aws-sf-m1-i1} shows an example. When the execution engine enters a new state of an instance, a corresponding event is captured for calculating $h$, as indicated in the figure, based on the trace of prior execution states and metadata including event IDs, timestamps, and the AWS region. Each event triggers the corresponding smart contract function in order to register hash values and metadata with the contract and to notify distributed clients through an event.

\subsubsection{Tracking and Analysis by Events and Instance Protocols}

From the client point-of-view, smart contract events are observed in order to store model, instance, and state events locally such that an instance protocol can be created in a database for tracking and analysis. Upon the notification by an event, the related functions retrieve the registered model, instance, or state data from AWS and verify it based on the smart contract data in the form of metadata, hash values, and related blockchain transactions. The verification follows the concept of blockchain-based attestation~\cite{harerDecentralizedAttestationDistribution2022a}, where validity is established through (1.) integrity, by re-calculation and comparison of hash values, (2.) time, by requiring matching event and model timestamps within a range, and (3.) client addresses linked to prior artifacts, by requiring addresses matching the owner of a model or instance as well as matching addresses for states of an instance and instances of a model. If validation is established, database records are created. 

\begin{figure}[htb]
  \centering
  \includegraphics[clip, trim=0.0cm 0.0cm 0.0cm 0.0cm, width=1.0\linewidth]{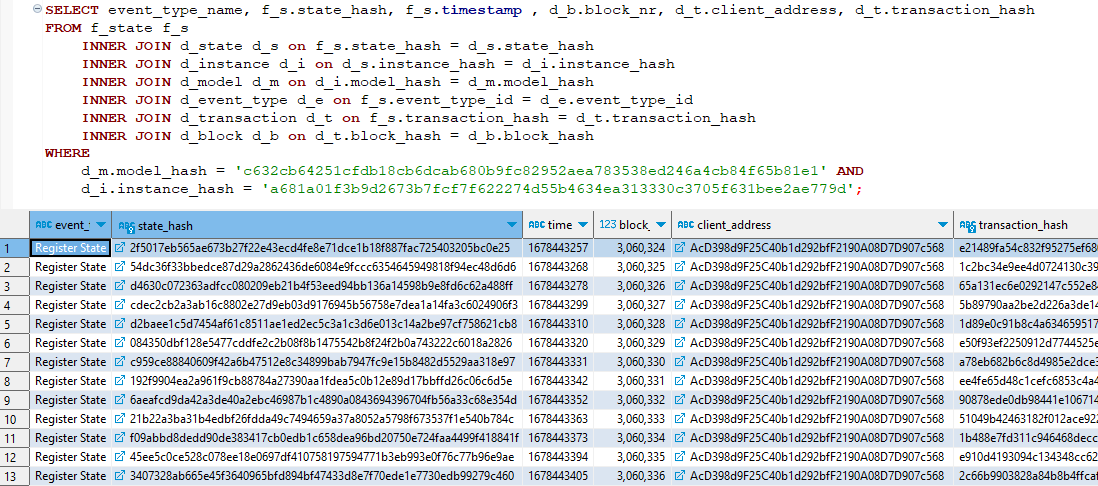}
  \caption{Instance protocol listing the 13 states of the instance in Figure~\ref{fig:aws-sf-m1-i1}. The protocol is retrieved by querying the multidimensional data model implemented in a relational PostgreSQL 15 database. }
  \label{fig:query1}
\end{figure}

\begin{figure}[htb]
  \flushleft
  \includegraphics[clip, trim=0.0cm 0.0cm 0.0cm 0.0cm, width=0.8\linewidth]{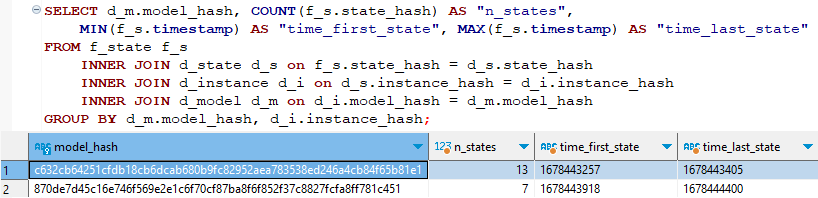}
  \caption{Comparison of the number of states between two instances by aggregation over model and instance dimensions.}
  \label{fig:query2}
\end{figure}

The database is implemented in the fashion of a data warehouse according to the data model to permit further analysis of past instances. Figure~\ref{fig:query1} shows an example query for retrieving an instance protocol of instance \textit{h = a681[...]779d} and model \textit{h = c632[...]81e1} as given before in Figures~\ref{fig:aws-sf-m1} and ~\ref{fig:aws-sf-m1-i1}, respectively. The states captured correspond to the states of type task state, i.e., states performing interactions with services for computation or data operations. Here, the second state shows the transformation of a data record in "Iteration \#0", the first of four concurrent iterations. The last state processed, by chance, is "Iteration \#2". In each iteration, the states named "Transform Data", "Update Warehouse", and "Dequeue" are reached, resulting in lines 2 to 13 of the instance protocol. The corresponding blockchain transactions between blocks 3060324 and 3060336 were made in the Ethereum Sepolia test network and can be seen there\footnote{\href{https://sepolia.etherscan.io/address/0xAcD398d9F25C40b1d292bfF2190A08D7D907c568}{https://sepolia.etherscan.io/address/0xAcD398d9F25C40b1d292bfF2190A08D7D90 7c568}}. As an example for analysis with aggregation along the dimensions, the number of states in the instance can be compared with, e.g., a second instance shown in Figure~\ref{fig:query2}. 

\section{Discussion and Outlook}
\label{discussion-outlook}

This paper investigates how decentralized applications can be executed on cloud platforms and blockchains such that distributed parties can observe execution behavior through the blockchain. For addressing the research question, instance tracking~\cite{harer_decentralized_2018,harerExecutableModelsInstance2022a} and attestation~\cite{harerDecentralizedAttestationDistribution2022a} concepts are applied based on an architecture for executing models on blockchains and cloud platforms~\cite{harerExecutableModelsInstance2022a}. 

As a first result, a metamodel describes the generalized concepts required for executing models on cloud platforms together with recording execution events using a smart contract and instance protocols. Secondly, a multidimensional data model realizes the concepts by permitting (a.) the recording of instance protocols locally for each distributed party in a verifiable manner, (b.) the tracking of states, instances, and models, and (c.) the analysis of past executions along multiple dimensions, e.g., specific clients, contracts, and time spans with optional data aggregations computing statistics such as the average number and duration of states. Thirdly, the implementation on the Ethereum blockchain and AWS with Step Function models confirms the feasibility of the approach. 

The approach combines cloud platforms and blockchains primarily for realizing scalability with the aim of decentralized coordination. While cloud platforms assume centralized control over the execution, they provide means to run applications distributed and abstract from concrete technical infrastructures, especially using serverless computing models such as AWS step functions. Through instance tracking the execution can gain transparency, however, the tracking does not address further aspects related to decentralization such as control and feedback mechanisms, or governance. Thus, decentralized applications can utilize blockchains with smart contracts for decentralized coordination aspects in combination with cloud platforms that support execution abstract from centralized technical infrastructure for gaining distribution and scalability.

%\section{Conclusion and Outlook}
%\label{discussion-outlook}

In effect, instance tracking allows for transparent execution behavior by distributing instance states in a verifiable manner to distributed parties. A consistent view of the tracking and analysis data is established, therefore, based on the locally stored instance protocols. Overall, the approach suggests tracking and analysis might be applied as part of decentralized applications on blockchains and cloud platforms. The transparency, distribution, and scalability properties gained on this basis might be utilized in data- and compute-intensive applications, e.g. involving IoT device data, machine learning, workflow automation, collaborations, or decentralized organizations. Future research will evaluate and address the integration of further decentralized execution aspects related to data- and compute-intensive applications.

\section*{Acknowledgment}

This work is supported by the Swiss National Science Foundation project Domain-Specific Conceptual Modeling for Distributed Ledger Technologies [196889].

\bibliographystyle{splncs04}
\bibliography{references}

\begin{thebibliography}{10}
\providecommand{\url}[1]{\texttt{#1}}
\providecommand{\urlprefix}{URL }
\providecommand{\doi}[1]{https://doi.org/#1}

\bibitem{amazon_2023}
Amazon: Aws step functions developer guide (2023),
  \url{https://docs.aws.amazon.com/step-functions/dg/}, accessed on 2023-03-12

\bibitem{bouragaTaxonomyBlockchainConsensus2021}
Bouraga, S.: A taxonomy of blockchain consensus protocols: {{A}} survey and
  classification framework. Expert Systems with Applications  \textbf{168}
  (2021). \doi{10.1016/j.eswa.2020.114384}

\bibitem{buterin_ethereum:_2013}
Buterin, V.: Ethereum: {{The Ultimate Smart Contract}} and {{Decentralized
  Application Platform}} (2013),
  \url{http://web.archive.org/web/20131228111141/http://vbuterin.com/ethereum.html},
  accessed on 2023-03-12

\bibitem{caiDecentralizedApplicationsBlockchainEmpowered2018}
Cai, W., Wang, Z., Ernst, J.B., Hong, Z., Feng, C., Leung, V.C.M.:
  Decentralized {{Applications}}: {{The Blockchain-Empowered Software System}}.
  IEEE Access  \textbf{6} (2018). \doi{10.1109/ACCESS.2018.2870644}

\bibitem{camundaCamundaBPMManual2022}
Camunda: The {{Camunda BPM Manual}}, bpm platform 7.18 (2022),
  \url{https://docs.camunda.org/manual/7.18/}, accessed on 2023-03-12

\bibitem{choudhuryAutoGenerationSmartContracts2018c}
Choudhury, O., Rudolph, N., Sylla, I., Fairoza, N., Das, A.:
  Auto-{{Generation}} of {{Smart Contracts}} from {{Domain-Specific
  Ontologies}} and {{Semantic Rules}}. In: 2018 {{IEEE International
  Conferences}} on Internet of Things, Green Computing and Communications,
  Cyber, Physical and Social Computing, Smart Data (2018).
  \doi{10.1109/Cybermatics_2018.2018.00183}

\bibitem{curty_blockchain_2022}
Curty, S., H{\"a}rer, F., Fill, H.G.: Blockchain application development using
  model-driven engineering and low-code platforms: {{A}} survey. In:
  Enterprise, Business-Process and Information Systems Modeling, {{EMMSAD}}
  2022. {Springer} (2022). \doi{10.1007/978-3-031-07475-2_14}

\bibitem{dotanSurveyBlockchainNetworking2021}
Dotan, M., Pignolet, Y.A., Schmid, S., Tochner, S., Zohar, A.: Survey on
  {{Blockchain Networking}}: {{Context}}, {{State-of-the-Art}}, {{Challenges}}.
  ACM Computing Surveys  \textbf{54}(5) (2021). \doi{10.1145/3453161}

\bibitem{fillKnowledgeBlockchainsApplying2018a}
Fill, H.G., H{\"a}rer, F.: Knowledge {{Blockchains}}: {{Applying Blockchain
  Technologies}} to {{Enterprise Modeling}}. In: 51st {{Hawaii International
  Conference}} on {{System Sciences}} ({{HICSS-51}}) (2018).
  \doi{10.24251/HICSS.2018.509}

\bibitem{garcia-garcia_using_2020}
{Garcia-Garcia}, J.A., {S{\'a}nchez-G{\'o}mez}, N., Lizcano, D., Escalona,
  M.J., Wojdy{\'n}ski, T.: Using {{Blockchain}} to {{Improve Collaborative
  Business Process Management}}: {{Systematic Literature Review}}. IEEE Access
  \textbf{8} (2020). \doi{10.1109/ACCESS.2020.3013911}

\bibitem{gosainConceptualMultidimensionalModeling2015}
Gosain, A., Singh, J.: Conceptual {{Multidimensional Modeling}} for {{Data
  Warehouses}}: {{A Survey}}. In: Proceedings of the 3rd {{International
  Conference}} on {{Frontiers}} of {{Intelligent Computing}}: {{Theory}} and
  {{Applications}} ({{FICTA}}) 2014. {Springer} (2015).
  \doi{10.1007/978-3-319-11933-5_33}

\bibitem{harer_decentralized_2018}
H{\"a}rer, F.: Decentralized {{Business Process Modeling}} and {{Instance
  Tracking Secured By}} a {{Blockchain}}. In: Proceedings of the 26th
  {{European Conference}} on {{Information Systems}} ({{ECIS}}). {AIS
  Electronic Library (AISeL)} (2018)

\bibitem{harerIntegrierteEntwicklungUnd2019}
H{\"a}rer, F.: Integrierte {{Entwicklung}} Und {{Ausf\"uhrung}} von
  {{Prozessen}} in Dezentralen {{Organisationen}}. {{Ein Vorschlag}} Auf
  {{Basis}} Der {{Blockchain}}. Dissertation. {University of Bamberg Press}
  (2019). \doi{10.20378/irbo-55721}

\bibitem{harerProcessModelingDecentralized2020}
H{\"a}rer, F.: Process {{Modeling}} in {{Decentralized Organizations Utilizing
  Blockchain Consensus}}. Enterprise Modelling and Information Systems
  Architectures (EMISAJ)  \textbf{15} (2020). \doi{10.18417/emisa.15.13}

\bibitem{harerExecutableModelsInstance2022a}
H{\"a}rer, F.: Executable {{Models}} and {{Instance Tracking}} for
  {{Decentralized Applications}} - {{Towards}} an {{Architecture Based}} on
  {{Blockchains}} and {{Cloud Platforms}}. In: Proceedings of the {{PoEM}} 2022
  {{Workshops}} and {{Models}} at {{Work}} Co-Located with {{Practice}} of
  {{Enterprise Modelling}} 2022. {{CEUR}}, vol.~3298 (2022)

\bibitem{harerInteroperabilityOpenPermissionless2022a}
H{\"a}rer, F.: Towards {{Interoperability}} of {{Open}} and {{Permissionless
  Blockchains}}: {{A Cross-Chain Query Language}}. In: 2022 {{IEEE
  International Conference}} on E-{{Business Engineering}} ({{ICEBE}}) (2022).
  \doi{10.1109/ICEBE55470.2022.00041}

\bibitem{harerDecentralizedAttestationConceptual2019b}
H{\"a}rer, F., Fill, H.G.: Decentralized {{Attestation}} of {{Conceptual Models
  Using}} the {{Ethereum Blockchain}}. In: 2019 {{IEEE}} 21st {{Conference}} on
  {{Business Informatics}} ({{CBI}}). {IEEE} (2019).
  \doi{10.1109/CBI.2019.00019}

\bibitem{harerDecentralizedAttestationDistribution2022a}
H{\"a}rer, F., Fill, H.G.: Decentralized attestation and distribution of
  information using blockchains and multi-protocol storage. IEEE Access
  \textbf{10} (2022). \doi{10.1109/ACCESS.2022.3150356}

\bibitem{hassanSurveyServerlessComputing2021}
Hassan, H.B., Barakat, S.A., Sarhan, Q.I.: Survey on serverless computing.
  Journal of Cloud Computing  \textbf{10}(1) (2021).
  \doi{10.1186/s13677-021-00253-7}

\bibitem{karagiannisMetamodelingApplicationAreas2008}
Karagiannis, D., Fill, H.G., H{\"o}fferer, P., Nemetz, M.: Metamodeling: {{Some
  Application Areas}} in {{Information Systems}}. In: Information {{Systems}}
  and E-{{Business Technologies}}. {Springer} (2008)

\bibitem{karagiannisDomainSpecificConceptualModeling2022}
Karagiannis, D., Lee, M., Hinkelmann, K., Utz, W. (eds.): Domain-{{Specific
  Conceptual Modeling}}: {{Concepts}}, {{Methods}} and {{ADOxx Tools}}.
  {Springer} (2022)

\bibitem{ladleifModelingEnforcingBlockchainBased2019b}
Ladleif, J., Weske, M., Weber, I.: Modeling and {{Enforcing Blockchain-Based
  Choreographies}}. In: Business {{Process Management}}: 17th {{International
  Conference}}, {{BPM}} 2019, {{Vienna}}, {{Austria}}, {{September}}
  1\textendash 6, 2019, {{Proceedings}}. {Springer} (2019).
  \doi{10.1007/978-3-030-26619-6_7}

\bibitem{liServerlessComputingSurvey2022}
Li, Z., Guo, L., Cheng, J., Chen, Q., He, B., Guo, M.: The {{Serverless
  Computing Survey}}: {{A Technical Primer}} for {{Design Architecture}}. ACM
  Computing Surveys  \textbf{54}(10s) (2022). \doi{10.1145/3508360}

\bibitem{lopez-pintadoCaterpillarBusinessProcess2019}
{L{\'o}pez-Pintado}, O., {Gar{\'c}{\i}a-Ba{\~n}uelos}, L., Dumas, M., Weber,
  I., Ponomarev, A.: Caterpillar: {{A}} business process execution engine on
  the {{Ethereum}} blockchain. Software: Practice and Experience
  \textbf{49}(7) (2019). \doi{10.1002/spe.2702}

\bibitem{mavridouDesigningSecureEthereum2018}
Mavridou, A., Laszka, A.: Designing {{Secure Ethereum Smart Contracts}}: {{A
  Finite State Machine Based Approach}}. In: Financial {{Cryptography}} and
  {{Data Security}}: 22nd {{International Conference}}, {{FC}} 2018. {Springer}
  (2018). \doi{10.1007/978-3-662-58387-6_28}

\bibitem{milaniModellingBlockchainbasedBusiness2021}
Milani, F., {Garc{\'i}a-Ba{\~n}uelos}, L., Filipova, S., Markovska, M.:
  Modelling blockchain-based business processes: A comparative analysis of
  {{BPMN}} vs {{CMMN}}. Business Process Management Journal  \textbf{27}(2)
  (2021). \doi{10.1108/BPMJ-06-2020-0263}

\bibitem{nadeemEvaluatingRankingCloud2022}
Nadeem, F.: Evaluating and {{Ranking Cloud IaaS}}, {{PaaS}} and {{SaaS Models
  Based}} on {{Functional}} and {{Non-Functional Key Performance Indicators}}.
  IEEE Access  \textbf{10} (2022). \doi{10.1109/ACCESS.2022.3182688}

\bibitem{nakamotoBitcoinPeertoPeerElectronic2008}
Nakamoto, S.: Bitcoin: {{A Peer-to-Peer Electronic Cash System}} (2008),
  \url{https://bitcoin.org/bitcoin.pdf}, accessed on 2023-03-12

\bibitem{nakamuraInterorganizationalBusinessProcesses2018}
Nakamura, H., Miyamoto, K., Kudo, M.: Inter-organizational {{Business Processes
  Managed}} by {{Blockchain}}. In: Web {{Information Systems Engineering}}
  \textendash{} {{WISE}} 2018. {Springer} (2018).
  \doi{10.1007/978-3-030-02922-7_1}

\bibitem{nistSecureHashStandard2015}
NIST: Secure {{Hash Standard}} ({{SHS}}). Tech. Rep. Federal Information
  Processing Standard (FIPS) 180-4, {U.S. Department of Commerce} (2015).
  \doi{10.6028/NIST.FIPS.180-4}

\bibitem{omgBusinessProcessModel2014}
{Object Management Group}: Business {{Process Model}} and {{Notation}}
  ({{BPMN}}) 2.0.2 (2014), \url{https://www.omg.org/spec/BPMN/2.0.2}, accessed
  on 2023-03-12

\bibitem{wangSoKDAGbasedBlockchain2023}
Wang, Q., Yu, J., Chen, S., Xiang, Y.: {{SoK}}: {{DAG-based Blockchain
  Systems}}. ACM Computing Surveys  \textbf{55}(12) (2023).
  \doi{10.1145/3576899}

\bibitem{wangDecentralizedAutonomousOrganizations2019}
Wang, S., Ding, W., Li, J., Yuan, Y., Ouyang, L., Wang, F.Y.: Decentralized
  {{Autonomous Organizations}}: {{Concept}}, {{Model}}, and {{Applications}}.
  IEEE Transactions on Computational Social Systems  \textbf{6}(5) (2019).
  \doi{10.1109/TCSS.2019.2938190}

\bibitem{weberUntrustedBusinessProcess2016}
Weber, I., Xu, X., Riveret, R., Governatori, G., Ponomarev, A., Mendling, J.:
  Untrusted {{Business Process Monitoring}} and {{Execution Using Blockchain}}.
  In: 14th {{International Conference}}, {{Business Process Management}}
  ({{BPM}} 2016). {Springer} (2016). \doi{10.1007/978-3-319-45348-4_19}

\bibitem{wuFirstLookBlockchainbased2021}
Wu, K., Ma, Y., Huang, G., Liu, X.: A first look at blockchain-based
  decentralized applications. Software: Practice and Experience
  \textbf{51}(10) (2021). \doi{10.1002/spe.2751}

\bibitem{ZhouScalability2020}
Zhou, Q., Huang, H., Zheng, Z., Bian, J.: Solutions to scalability of
  blockchain: {{A}} survey. IEEE Access  \textbf{8} (2020).
  \doi{10.1109/ACCESS.2020.2967218}

\end{thebibliography}

\end{document}